\documentstyle[12pt,aasms4]{article}

\newcommand {\ie} {{\it  i.e.}}

\newcommand {\go} {\mathrel{\hbox{\rlap{\lower.55ex \hbox {$\sim$}}
        \kern-.3em \raise.4ex \hbox{$>$}}}}
\newcommand {\lo} {\mathrel{\hbox{\rlap{\lower.55ex \hbox {$\sim$}}
        \kern-.3em \raise.4ex \hbox{$<$}}}}
\newcommand {\be} {\begin{equation}}
\newcommand {\ee} {\end{equation}}

\def\half{\ifmmode {1\over 2} \else ${1\over 2}$\fi} 
\def\tor{\ifmmode {3\over r} \else ${3\over r}$\fi} 
\def\dwds{\ifmmode \partial W/\partial \Sigma \else $\partial W/
\partial \Sigma$\fi} 
\def\dwdr{\ifmmode \partial W/\partial r \else $\partial W/
\partial r$\fi} 
\def\dsdr{\ifmmode \partial \Sigma/\partial r \else $\partial \Sigma/
\partial r$\fi} 
\def\SigG{\ifmmode \Sigma_{GPD} \else $\Sigma_{GPD}$\fi} 
\def\SigR{\ifmmode \Sigma_{RPD} \else $\Sigma_{RPD}$\fi} 
\def\Msun{\ifmmode M_\odot \else $M_\odot$\fi} 
\def\Mdot{\ifmmode \dot M \else $\dot M$\fi} 
\def\rms{\ifmmode r_{ms} \else $r_{ms}$\fi} 
\def\rin{\ifmmode r_{in} \else $r_{in}$\fi} 
\def\kes{\ifmmode \kappa_{es} \else $\kappa_{es}$\fi}
\def\fK{\ifmmode \nu_K \else $\nu_K$\fi} 

\begin{document}

\title{COMMENT ON VISCOUS STABILITY OF RELATIVISTIC KEPLERIAN
ACCRETION DISKS}
\author{Rafa{\l} Moderski and Bo\.zena Czerny}
\affil{Nicolaus Copernicus Astronomical Center, 00-716 Warsaw,
Bartycka 18, POLAND} 

\lefthead{Czerny \& Moderski}
\righthead{Stability of Relativistic Disks}

\begin{abstract}
  Recently Ghosh~(1998) reported a new regime of instability in
  Keplerian accretion disks which is caused by relativistic
  effects. This instability appears in the gas pressure dominated
  region when all relativistic corrections to the disk structure
  equations are taken into account. We show that he uses the stability
  criterion in completely wrong way leading to inappropriate
  conclusions. We perform a standard stability analysis to show that
  no unstable region can be found when the relativistic disk is gas
  pressure dominated.
\end{abstract}


\section{Introduction}

Stability analysis of Shakura-Sunyaev $\alpha$ disk (Shakura \&
Sunyaev~1973) was performed by many authors (Lightman \& Eardley 1974;
Lightman 1974; Shakura \& Sunyaev 1976; Piran 1978) and yield the well
known conclusions: the disk is stable for gas pressure dominated (GPD)
region and becomes viscously and thermally unstable for radiation
pressure dominated (RPD) region. While the general relativity
generalization of the Shakura \& Sunyaev model were available (Novikow
\& Thorne~1973) the stability studies were presented only for the
Newtonian case.

Recently Ghosh~(1998) reported a new regime of instability for
$\alpha$ disks when general relativity effects are taken into
account. He found that even in GPD disks there exist a region not far
from the inner edge of the disk which is viscously unstable. This
might have an enormous influence on the models of active galactic
nuclei and galactic low-mass X-ray binaries.

In this paper we repeat a stability analysis studied by Lightman \&
Eardley~(1974) but in a relativistic case. We obtain the same results
as for the Newtonian case \ie, disk is viscously stable for gas
pressure dominated region, but becomes viscously unstable in the
region where radiation pressure dominates. Thus we do not confirm the
results obtained by Ghosh~(1998). We discuss an error made by
Ghosh~(1998) leading to wrong conclusions.

\section{Disk structure}

Surface density evolution is described by the equation (Ghosh~1998)
\be
{\cal A}^{\half}{\cal D}^{-\half}{\partial\Sigma\over\partial t} =
{\cal D}^{\half}{1\over r}{\partial\over\partial r}\left[ {{\cal
D}^{-\half}\over (\partial\ell/\partial r)} {\partial\over\partial
r}\left({\cal BC}^{-1}{\cal D}r^2W\right)
\right]\,,
\label{Sigma}
\ee
where $W$ is the vertically integrated viscous stress, $\ell$ is the
specific angular momentum, and $\cal A$, $\cal B$, $\cal C$, $\cal D$
are relativistic corrections (Novikov \& Thorne 1973)
\be
  {\cal A}\equiv 1+{A^2 \over \tilde r^2}+{2 A^2 \over \tilde r^3},\quad 
  {\cal B}\equiv 1+{A \over \tilde r^{3/2}},\quad
  {\cal C}\equiv 1-{3\over \tilde r}+{2 A \over \tilde r^{3/2}}, \quad
  {\cal D}\equiv 1-{2\over \tilde r}+{A^2 \over \tilde r^2}\,.
\ee
Here $A \equiv cJ/GM^2$ is the dimensionless spin of the central accreting 
object, and $\tilde r \equiv r c^2/GM$ is the dimensionless distance.

Equations of disk structure under the assumption of thermal
equilibrium yield following expressions for the relation between the
surface density of the disk and the viscous stress
\be
\SigR = \left({16 c^2 \over 9 \alpha \kes^2}\right)W^{-1}
{\cal CKD}^{-2}\,,
\label{SigR}
\ee
\be
\SigG = \left({m_p\over k \alpha}\right)^{4/5} \left({b c \over 
18\kes}\right)^{1/5} \left( {r^3 \over G M} \right)^{1/10}W^{3/5}
{\cal C}^{1/5}{\cal D}^{-1/5}\,,
\label{SigG}
\ee
where, ${\cal K} \equiv 1 - {4 A \over \tilde r^{3/2}} + {3 A^2 \over
\tilde r^2}$ is relativistic correction introduced by Riffert \&
Herold~(1995); $\SigG$ and $\SigR$ are surface densities in the gas
pressure and radiation pressure dominated regimes respectively.

\section{Viscous stability analysis}

Newtonian version of Equations (\ref{Sigma}) and (\ref{SigR}) was used
by Lightman \& Eardley~(1974; see also Lightman~1974) to study the
stability of the Keplerian disk. The assumption of thermal equilibrium
was proved to be justified by Shakura \& Sunyaev~(1976) because in the
limit of wavelengths of the radial perturbation much longer than the
disk thickness the viscous instability is clearly decoupled from the
thermal instability. As was correctly noticed by Ghosh~(1998), the
basic stability criterion is the sign of the partial derivative,
\dwds, and remains the same for both relativistic and non relativistic
disks.

The point is that in order to determine the stability of the disk we
should compute this partial derivative at a given radius directly from
Equations (\ref{SigR}) and (\ref{SigG}). It corresponds to the linear
perturbation of the surface density distribution in
Equation~(\ref{Sigma}) under the assumption of thermal equilibrium
leading to known relation $W(\Sigma)$ at a given radius.  In that case
the perturbation $\delta W$ of $W$ is expressed through
\be
 \delta W = (\dwds)_r \delta \Sigma, 
\ee
and in the next step $\delta \Sigma$ is decomposed into linear waves
according to standard local analysis. Spatial derivatives of
unperturbed stress and surface density are neglected in this process.

Thus for radiation dominated disk we have
\be
W \propto (\alpha \Sigma)^{-1} {\cal CKD}^{-2}
\label{WR}
\ee
>From (\ref{WR}) we can calculate viscous stress derivative
\be
\left (\dwds \right )_{RPD} \propto - \alpha^{-1} \Sigma^{-2} {\cal
CKD}^{-2} < 0
\ee
which immediately shows that the disk is viscously unstable in the
whole radiation pressure dominated region. This is a relativistic
generalization of the well known result obtained first by Lightman \&
Eardley~(1974) exactly in the same way.

For gas pressure dominated disk from (\ref{SigG}) we have
\be
\left (\dwds \right )_{GPD} \propto \Sigma^{2/3} r^{-11/6} {\cal
C}^{-1/3} {\cal D}^{1/3} > 0
\ee
which shows that gas pressure dominated disk is stable. No evidence is
found for any unstable region, in opposite to the claim of
Ghosh~(1998).

The conclusion of Ghosh~(1998) was based on the computation of
entirely different partial derivative, namely $(\dwds)_{\dot M}$, or
equivalently, the expression $(\dwdr)/(\dsdr)$ which has no
straightforward relation to $(\dwds)_r$.

To show this clearly we first replot the Figure~1 of Ghosh~(1998)
showing the radial dependence of the stress and the surface density in
a stationary solution for a single value of accretion rate. The
loop-like character of this plot simply reflects the fact that the
stress and the surface density are zero at the marginally stable orbit
(upper branch) and they again tend to zero for radius approaching
infinity (lower branch). In this Figure the region where
$(\dwdr)/(\dsdr)<0$ corresponds to radii between $\tilde r \simeq 14$
and $\tilde r \simeq 22$.

Since Ghosh~(1998) claimed that the negative part of the slope
indicate the unstable region we show the correct stability plot for
the two radii from this range. A single such plot, as explained above,
show the dependence of the stress vs. surface density for a fixed
radius but variable accretion rate. Upper and lower part of a curve is
well approximated by Equation~(\ref{SigR}) and Equation~(\ref{SigG}),
correspondingly, and more general approach to gas/radiation pressure
contribution matches smoothly those two asymptotic solutions giving a
single curve. The slop of this curve determines the disk viscous
stability. The accretion rate adopted in Figure~1 is marked in
Figure~2 with crosses. It is well within a stable region, as claimed
in all papers prior to Ghosh~(1998).

To confirm our analysis we perform numerical simulations of surface
density evolution according to Equation~(\ref{Sigma}) for both
radiation pressure and gas pressure dominated disks. The results are
presented on Figure~3 and 4 respectively. We can see that for
radiation pressure dominated disk the initial small perturbation grows
leading to total disruption of the disk, while for gas pressure
dominated disk initial perturbation is smoothed out to stationary
configuration.

\section{Conclusions}

We study the viscous stability of relativistic, Keplerian accretion
disk in the thermal equilibrium. We confirm the results previously
obtained in the Newtonian case \ie, the disk is stable in gas pressure
dominated regime and becomes viscously unstable in radiation pressure
dominated regime. We show that instability recently reported by
Ghosh~(1998) was caused by improper use of the stability criterion for
accretion disk, and does not appear if the correct approach is
applied. We confirm our results by performing numerical simulations of
surface density evolution which completely agree with our analytical
analysis.

\hskip 1in

\acknowledgements

This work was supported by Polish KBN grants 2P03D00410 and
2P03D00415. RM also acknowledge support from the Foundation for Polish
Science.

\clearpage

\begin{figure}

\centerline{\bf FIGURE CAPTIONS}

\caption {Integrated viscous stress $W$ versus surface density
$\Sigma$ obtained in the same way like Ghosh~(1998). The accretion
rate is fixed ($\dot m = 0.01$) and the quantity used to parameterize
the plot along the curve is the distance from the central object. This
relation shows the ``unstable region'' between $\tilde r \sim 14$
(maximum of $W$) and $\tilde r \sim 22$ (maximum of $\Sigma$)}

\caption {Proper relation between integrated viscous stress and
surface density usually used for stability analysis. This time, for a
given curve, the radius is fixed and accretion rate is used to
parameterize the plot. Two curves are shown: for $\tilde r = 15$ and
$\tilde r = 21$. Accretion rate from the Fig.~1 is marked by big
crosses. There is no evidence for instability in these points.}

\caption {Evolution of the surface density of the radiation pressure
dominated disk. Small initial perturbation grows and finally disrupt
the disk.}

\caption{Same as Figure~3 but for gas pressure dominated disk. Initial
perturbation is placed at $\tilde r = 18$ (inside the ``unstable
region'' postulated by Ghosh~(1998)). Perturbation is smoothed out
showing no evidence for instability.}

\end{figure}


\begin{thebibliography}{}

\bibitem[]{}
Ghosh, P. 1998, \apj, 506, L113

\bibitem[]{}
Lightman, A. P. 1974, \apj, 194, 429

\bibitem[]{}
Lightman, A. P. \& Eardley, D. M. 1974, \apj, 187, L1

\bibitem[]{}
Novikov, I. D. \& Thorne, K. S. 1973, in Black Holes, ed.
C. DeWitt \& B. S. DeWitt (New York: Gordon \& Breach), 
343 (NT73)

\bibitem[]{}
Piran, T. 1978, \apj, 221, 652

\bibitem[]{}
Riffert, H. \& Herold, H. 1995, \apj, 450, 508

\bibitem[]{}
Shakura, N.I. \& Sunyaev, R.A. 1973, \aap, 24, 337

\bibitem[]{}
Shakura, N.I. \& Sunyaev, R.A. 1976, \mnras, 175, 613  

\end{thebibliography}
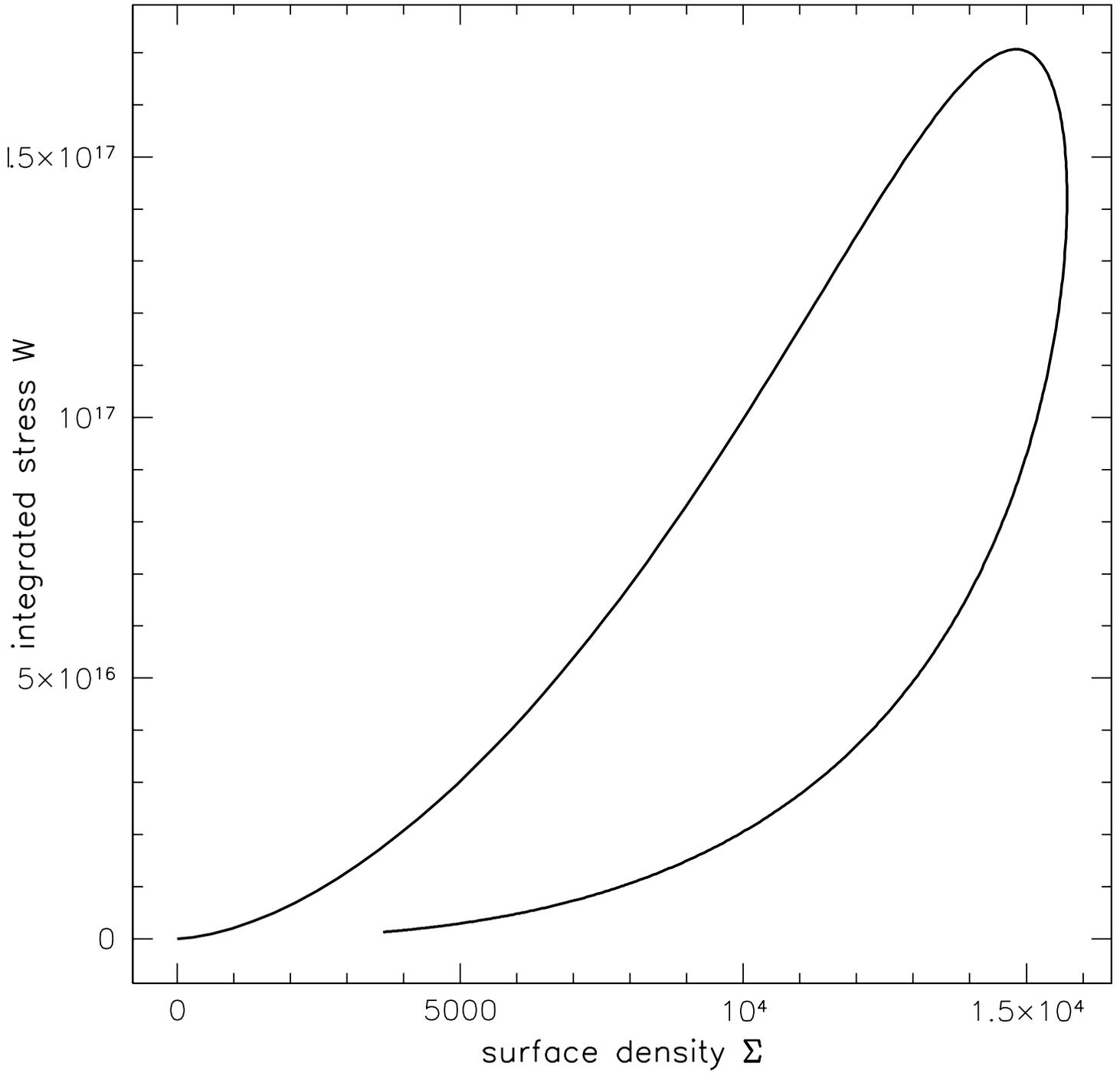
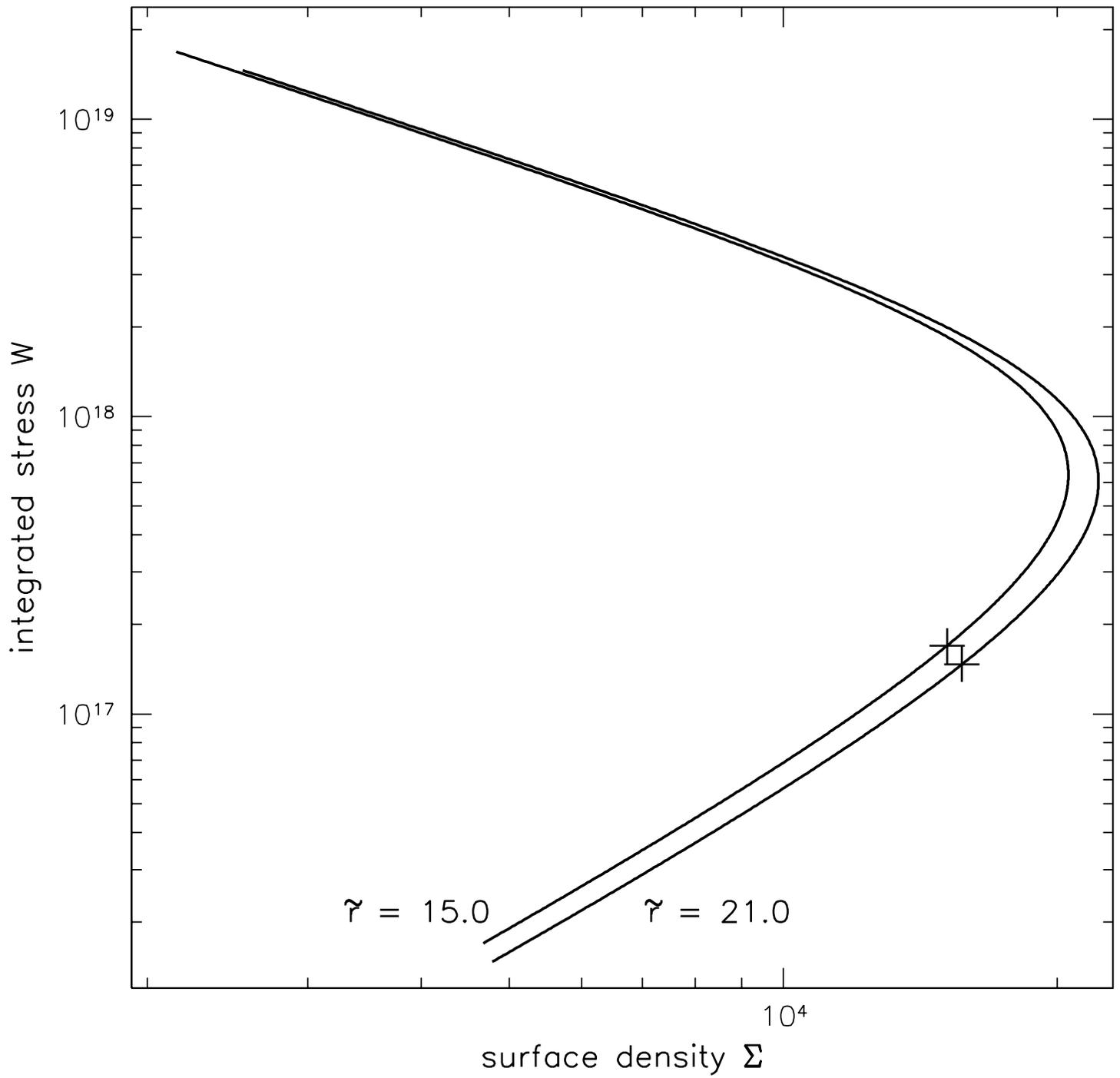
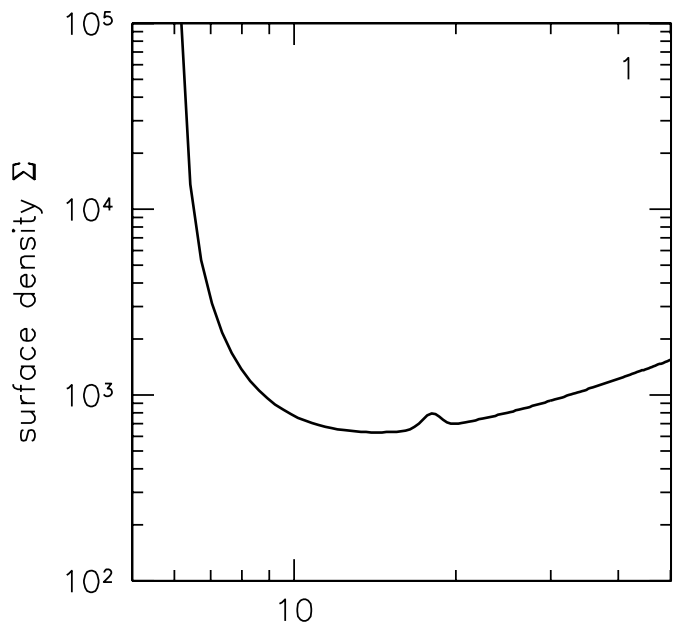
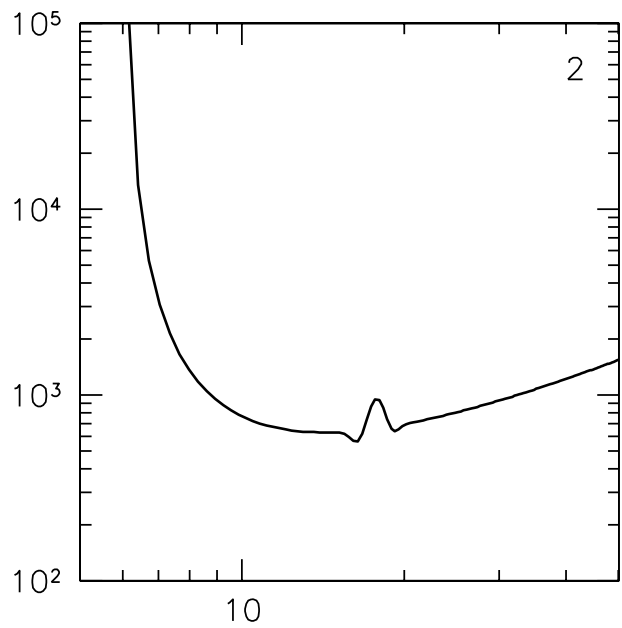
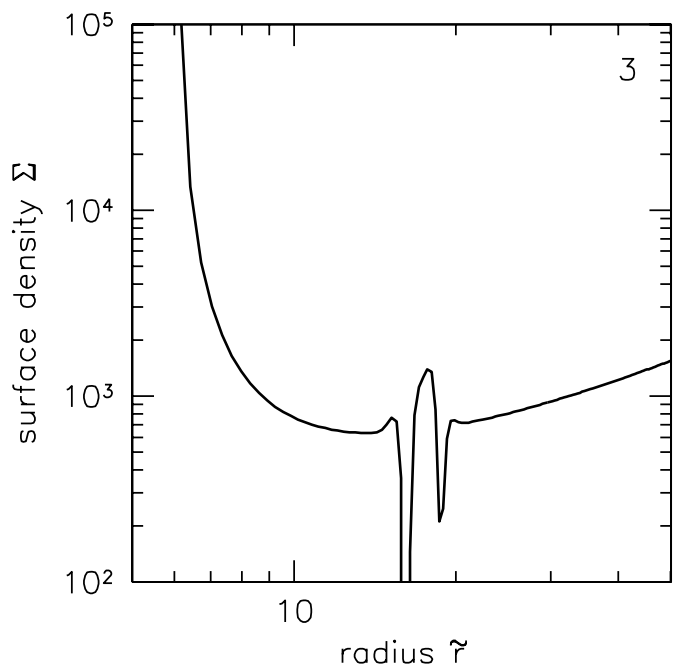
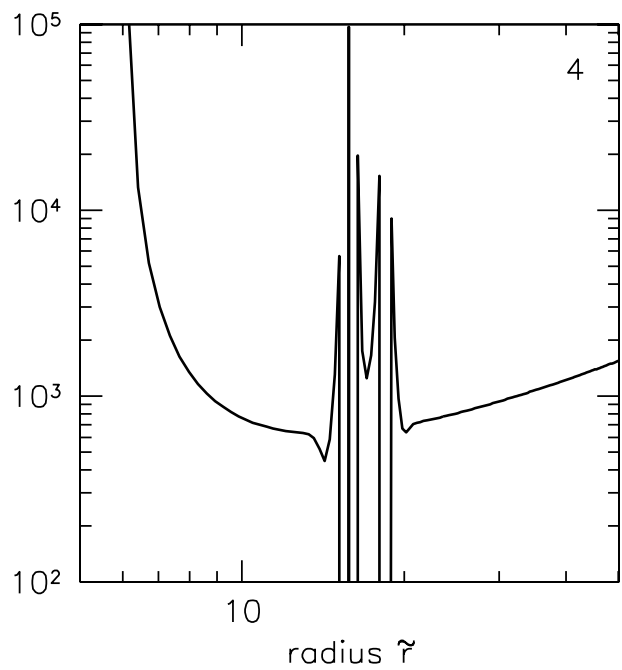
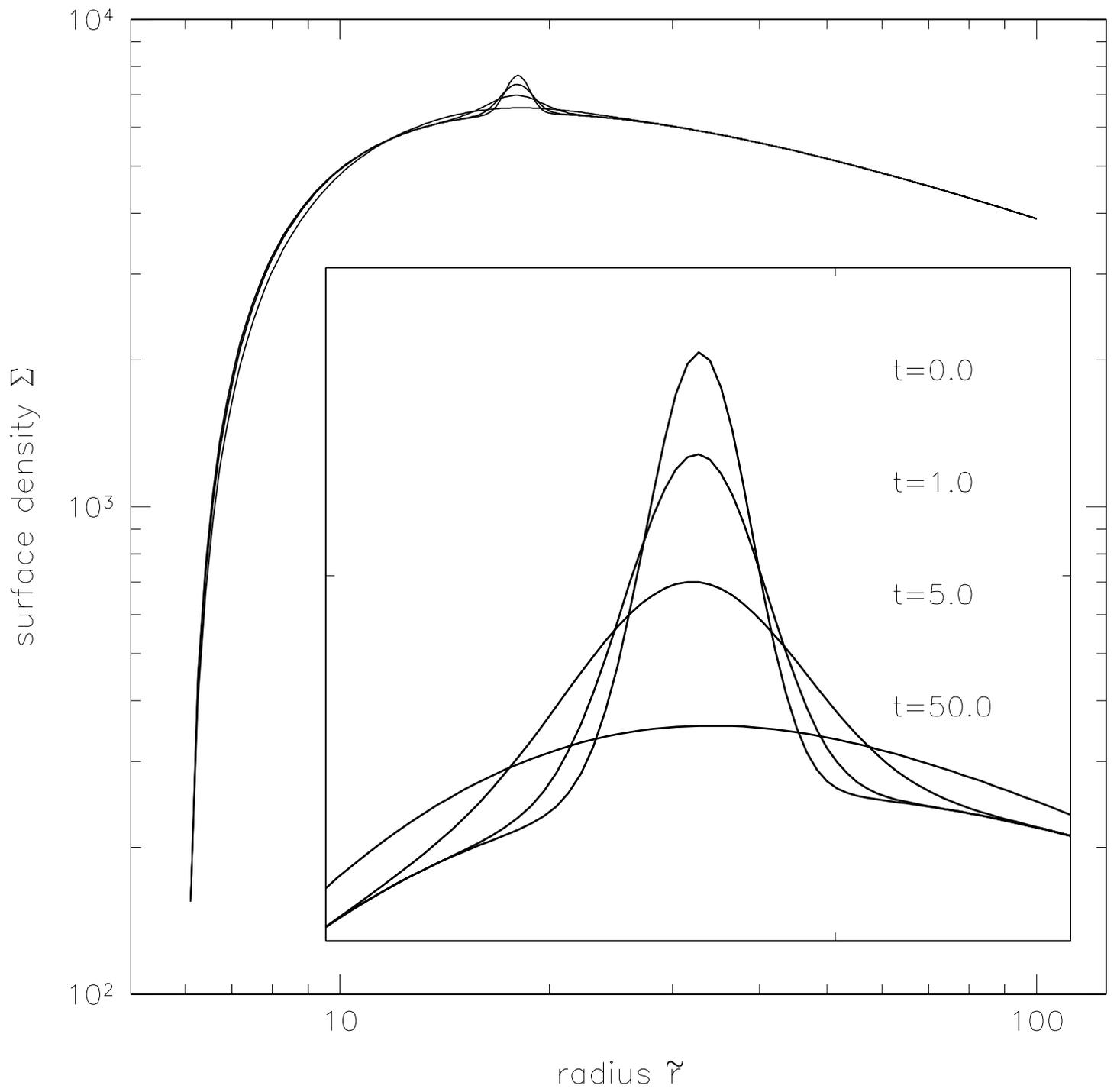
\end{document}